\documentclass{article}

\begin{document}

\author{J.J. Rosales$ \footnote{E-mail:
rosales@salamanca.ugto.mx}$\\
Facultad de Ingenier\'{\i}a Mec\'anica, El\'ectrica y
Electr\'onica. \\
Campus FIMEE, Universidad de Gto.\\
Carretera Salamanca-Valle de Santiago, km. 3.5 + 1.8 km.\\
Comunidad de Palo Blanco, Salamanca Gto. M\'exico.\\
\\
V.I. Tkach\\
Department of Physics and Astronomy\\
Northwestern University\\
Evanston, IL 60208-3112, USA}
\title{Supersymmetric Barotropic FRW Model and Dark Energy}
\maketitle

{\bf Abstract:} Using the superfield approach we construct the
$n=2$ supersymmetric lagrangian for the FRW Universe with
barotropic perfect fluid as matter field. The obtained
supersymmetric algebra allowed us to take the square root of the
Wheeler-DeWitt equation and solve the corresponding quantum
constraint. This model leads to the relation between the vacuum
energy density and the energy density of the dust matter.
\\


\begin{center}
{\bf Introduction}
\end{center}

Einstein's theory of general relativity is by far the most
attractive classical theory of gravity today. By describing the
gravitational field in terms of the structure of space-time,
Einstein effectively equated the study of gravity with the study
of geometry. In general relativity, space-time is a
$4$-dimensional manifold with Lorentzian metric $g_{\mu\nu}$ whose
curvature measure the strength of the gravitational field. Given a
matter distribution described by a stress-energy tensor
$T_{\mu\nu}$, the curvature of the metric is determined by
Einstein equations $G_{\mu\nu} = 8 \pi G T_{\mu\nu}$. This tensor
equation completely describes the classical theory.
\\

The first attempts to develop a quantum theory of gravity are
almost as old as quantum field theory itself. With the development
by Dirac$^{1,2}$ of a consistent treatment for constrained
Hamiltonian systems, the way was paved for the canonical
formulation of a quantum theory of gravity. Today, the use of the
canonical formalism in the  reduction of the Einstein
gravitational action to Hamiltonian form is well known $^{3,4,5}$.
However, the passage from the classical to the quantum theory
using the substitution of dynamical variables by operators and
Poisson brackets by commutators is complicated by the problem of
operator ordering $^{6,7,8}$, so that one is left with the choice
of either abandoning the canonical approach or studying simplified
models called minisuperspace. The Minisuperspaces are useful toy
models for canonical quantum gravity, because they capture many of
the essential features of general relativity and are at the same
time free of technical difficulties associated with the presence
of an infinite number of degrees of freedom. The Bianchi
cosmologies are the prime example. As is well known, the equation
that governs the quantum behavior of these models is the
Wheeler-DeWitt equation, which results in a quadratic Hamiltonian
leading to an equation of the Klein-Gordon type. The introduction
of supersymmetric minisuperspace models has led to the definition
and study of linear "square root" equations defining the quantum
evolution of the Universe. To achieve these Dirac-Type equations
one can make use the fact that supergravity provides a natural
square root of gravity $^{9-21}$ or supersymmetrize the models
$^{22-27}$.\\

Some time ago we have used the superfield formulation to
investigate supersymmetric cosmological models $^{28,29}$. In the
previous works $^{30-34}$ it was shown that the spatially
homogeneous part of the fields in the supergravity theory
preserves the invariance under the local time $n = 2$
supersymmetry. This supersymmetry is a subgroup of the four
dimensional space-time supersymmetry of the supergravity theory.
This local supersymmetry procedure has the advantage that, by
defining the superfields on superspace, all the component fields
in a supermultiplet can be manipulated simultaneously in a manner
that automatically preserves supersymmetry. Besides, the fermionic
fields are obtained in a clear manner as the supersymmetric
partners of the cosmological bosonic variables.
\\

More recently, using the superfield formulation the canonical
procedure quantization for a closed FRW cosmological model filled
with pressureless matter (dust) content and the corresponding
superpartner was reported $^{30-34}$. We have obtained the
quantization for the energy-like parameter, and it was shown, that
this energy is associated with the mass parameter quantization,
and that such type of Universe has a quantized masses of the order
of the Planck mass.
\\

In the present work we are interested in the construction of the
$n=2$ supersymmetric lagrangian for the FRW Universe with
barotropic perfect fluid as matter field including the
cosmological constant. The obtained supersymmetric algebra allowed
us to take the square root of the Wheeler-DeWitt equation and
solve the corresponding quantum constraint.
\begin{center}
{\bf Classical Action}
\end{center}
The classical action for a pure gravity system and the
corresponding term of matter content, perfect fluid with a
constant equation of state parameter $\gamma$; $p = \gamma \rho$,
and the cosmological term is $^{30-34}$
\begin{equation}
S = \int\Big[ -\frac{c^2 R}{2N \tilde G}\Big(\frac{dR}{dt}\Big)^2
+ \frac{Nkc^4}{2\tilde G}R + \frac{Nc^4 \Lambda}{6 \tilde G} R^3 +
NM_{\gamma}c^2 R^{-3\gamma}\Big] dt. \label{2}
\end{equation}
where $c$ is the velocity of light in vacuum, $ \tilde G = \frac{8
\pi G}{6}$ where $G$ is the Newtonian gravitational constant; $k =
1,0, -1$ stands for spherical, plane or hyperspherical three
space; $N(t), R(t)$ are the lapse function and the scale factor,
respectively; $M_\gamma$ is the mass by unit
${\rm length}^{-\gamma}$.\\

The purpose of this work is the supersymmetrization of the full
action (\ref{2}) using the superfield approach. The action
(\ref{2}) is invariant under the time reparametrization
\begin{equation}
t^{\prime} \to t + a(t), \label{3}
\end{equation}
if the transformations of $R(t)$ and $N(t)$ are defined as
\begin{equation}
\delta R = a {\dot R}, \qquad \delta N = (aN)^{.}\label{4}
\end{equation}
The variation with respect to $R(t)$ and $N(t)$ lead to the
classical equation for the scale factor $R(t)$ and the constraint,
which generates the local reparametrization of $R(t)$ and $N(t)$.
This constraint leads to the Wheeler-DeWitt equation in quantum
cosmology.
\\

In order to obtain the corresponding supersymmetric action for
(\ref{2}), we follow the superfield approach. For this, we extend
the transformation of time reparametrization (\ref{3}) to the $n=
2$ local supersymmetry of time $(t, \eta, \bar\eta)$. Then, we
have the following local supersymmetric transformation
\begin{eqnarray} \delta{t}  &  = & a(t) +
\frac{i}{2}[\eta\beta^{\prime}(t) + \bar
\eta{\bar\beta^{\prime}(t)}],\nonumber\\
\delta\eta &  = & \frac{1}{2}\bar\beta^{\prime}(t)+ \frac{1}{2}
[\dot a(t) + ib(t)]\eta +
\frac{i}{2}\dot{\bar\beta}^{\prime}(t)\eta\bar\eta
,\label{5}\\
\delta{\bar\eta}  &  = & \frac{1}{2}\beta^{\prime}(t) +
\frac{1}{2}[\dot a(t) - ib(t)]\bar\eta- \frac{i}{2}
\dot\beta^{\prime}(t) \eta\bar\eta,\nonumber
\end{eqnarray}
where $\eta$ is a complex odd parameter ($\eta$ odd ``time''
coordinates), $\beta^{\prime}(t) = N^{-1/2}\beta(t)$ is the
Grassmann complex parameter of the local ``small'' $n=2$
supersymmetry (SUSY) transformation, and $b(t)$ is the parameter
of local $U(1)$ rotations of the complex $\eta$.
\\

For the closed $(k = 1)$ and plane $(k = 0)$ FRW action we propose
the following superfield generalization of the action (\ref{2}),
invariant under the $n = 2$ local supersymmetric transformation
(\ref{5})
\begin{eqnarray}
S_{susy} &=& \int \Big[ -\frac{c^2}{2 \tilde G} {I\!\!N}^{-1}
{I\!\!R}D_{\bar\eta}{I\!\!R} D_\eta{I\!\!R} + \frac{c^3 {\sqrt
k}}{2 \tilde G} {I\!\!R}^2 + \frac{c^3
\Lambda^{1/2}}{3\sqrt{3}\tilde G}{I\!\!R}^3 -\nonumber\\
&-& \frac{2 \sqrt{2}M^{1/2}_\gamma}{(3 - 3\gamma) \tilde G^{1/2}}
{I\!\!R}^{\frac{3-3\gamma}{2}}\Big]d{\eta} d{\bar\eta}dt,
\label{6}
\end{eqnarray}
where
\begin{equation}
D_{\eta} = \frac{\partial }{\partial \eta} + i\bar\eta
\frac{\partial }{\partial t}, \qquad D_{\bar\eta} =
-\frac{\partial }{\partial \bar\eta} - i\eta \frac{\partial
}{\partial t}, \label{7}
\end{equation}
are the supercovariant derivatives of the global "small"
supersymmetry of the generalized parameter corresponding to $t$.
The local supercovariant derivatives have the form ${\tilde
D}_{\eta} = {I\!\!N}^{-1/2} D_\eta$, ${\tilde D}_{\bar\eta} =
{I\!\!N}^{-1/2} D_{\bar\eta}$, and ${I\!\!R}(t,\eta, \bar\eta),
{I\!\!N}(t,\eta, \bar\eta)$ are superfields. The supersymmetric
action for $\Lambda = 0$, $\gamma = 0$ was reported in \cite{17}.

The Taylor series expansion for the superfields
${I\!\!N}(t,\eta,\bar\eta)$ and ${I\!\!R}(t,\eta,\bar\eta)$ are
the following
\begin{eqnarray}
{I\!\! N}(t,\eta,\bar\eta)&=& N(t) + i\eta\bar\psi^{\prime}(t) +
i\bar
\eta\psi^{\prime}(t) + V^{\prime}(t)\eta\bar\eta,\label{8}\\
{I\!\! R}(t,\eta,\bar\eta) &=& R(t) +
i\eta\bar\lambda^{\prime}(t)+
i\bar\eta\lambda^{\prime}(t) + B^{\prime}(t) \eta\bar\eta.\label{9}%
\end{eqnarray}
In the expressions (\ref{8}) and (\ref{9}) we have introduced the
redefinitions $\psi^{\prime}(t) = N^{1/2}\psi(t)$, $V^{\prime} =
N(t)V(t) + \bar\psi(t) \psi(t)$, $\lambda^{\prime} = \frac{\tilde
G^{1/2}N^{1/2}}{cR^{1/2}} \lambda$ and $B^{\prime} = \frac{\tilde
G^{1/2}}{c}NB + \frac{\tilde G^{1/2}}{2c R^{1/2}}(\bar\psi \lambda
- \psi \bar\lambda)$. The components of the superfield
${I\!\!N}(t, \eta, \bar\eta)$ are gauge fields of the
one-dimensional $n=2$ extended supergravity. $N(t)$ is the
einbein, $\psi(t), \bar\psi(t)$ are the complex gravitino fields,
and $V(t)$ is the $U(1)$ gauge field. The component $B(t)$ in
(\ref{9}) is an auxiliary degree of freedom (non-dynamical
variable), and $\lambda, \bar\lambda$ are the fermion partners of
the scale factor $R(t)$. Thus, we can rewrite the action (\ref{6})
in its component form
\begin{eqnarray}
S_{susy} &=& \int \left\{- \frac{c^2 R (DR)^2}{2N \tilde G} +
\frac{i}{2}(\bar\lambda D\lambda - D\bar\lambda \lambda) -
\frac{NR}{2}B^2 - \frac{N \tilde G^{1/2} B}{2cR}\bar\lambda
\lambda + \right. \nonumber\\
&& \left. + \frac{c^2 \sqrt{k} RN}{\tilde G^{1/2}} B + \frac{c^2
\sqrt{k} R^{1/2}}{2\tilde G^{1/2}}(\bar\psi \lambda - \psi
\bar\lambda) + \frac{cN \sqrt{k}}{R}\bar\lambda \lambda + \right.
\label{10}\\
&& \left. + \frac{c^2 \Lambda^{1/2}}{\sqrt{3} \tilde G^{1/2}}
NR^2B + \frac{c^2 \Lambda^{1/2} R^{3/2}}{2\sqrt{3} \tilde G^{1/2}}
(\bar\psi \lambda - \psi \bar\lambda) + \frac{2 c \Lambda^{1/2}
N}{\sqrt{3}} \bar\lambda \lambda - \right. \nonumber\\ && \left. -
\sqrt{2}cM^{1/2}_{\gamma}N R^{\frac{1 - 3\gamma}{2}} B -
\frac{\sqrt{2}}{2} c
M_{\gamma}^{1/2}R^{-\frac{3\gamma}{2}}(\bar\psi \lambda - \psi
\bar\lambda) - \right. \nonumber\\
&& \left. - \sqrt{2}(1 - 3\gamma)\tilde G^{1/2}M_{\gamma}^{1/2}N
R^{\frac{-3 - 3\gamma}{2}}\bar\lambda \lambda \right
\}dt.\nonumber
\end{eqnarray}
So, the lagrangian for the auxiliary field has the form
\begin{eqnarray}
L_B &=& -\frac{N R}{2} B^2 - \frac{N \tilde G^{1/2} B}{2 cR}
\bar\lambda \lambda  + \frac{c^2 \sqrt{k} RN}{\tilde G^{1/2}}B +
\frac{c^2 \Lambda^{1/2} N R^2 }{\sqrt{3} \tilde G^{1/2}}B -
\nonumber\\
&& - \sqrt{2} c M_{\gamma}^{1/2} NR^{\frac{1 - 3\gamma}{2}} B.
\label{11}
\end{eqnarray}
From the expression (\ref{11}) we can obtain the equation for the
auxiliary field varying the Lagrangian with respect to $B$
\begin{equation}
B = \frac{c^2 \sqrt{k}}{ \tilde G^{1/2}} -\frac{ \tilde
G^{1/2}}{2cR^2}\bar\lambda \lambda + \frac{c^2 \Lambda^{1/2} R}{
\sqrt{3} \tilde G^{1/2}} - \sqrt{2} c M_{\gamma}^{1/2}
R^{\frac{-3\gamma - 1}{2}}. \label{12}
\end{equation}
Then, putting the expression (\ref{12}) in (\ref{10}) we have the
following supersymmetric action
\begin{eqnarray}
S_{susy} &=& \int \left\{- \frac{c^2 R (DR)^2}{2N \tilde G} +
\frac{c^4 NkR}{2\tilde G} + \frac{c^4 N \Lambda R^3}{6 \tilde G} +
Nc^2M_\gamma R^{-3\gamma} + \right.
\nonumber\\
&& \left. + \frac{c^4 \sqrt{k} \Lambda^{1/2} R^2}{\sqrt{3} \tilde
G} -  \frac{\sqrt{2k}c^3}{\tilde G^{1/2}}
M_{\gamma}^{1/2}R^{\frac{1 - 3\gamma}{2}} -
\frac{\sqrt{2}c^3\Lambda^{1/2} M_\gamma^{1/2}}{\sqrt{3} \tilde
G^{1/2}} R^{\frac{3 - 3 \gamma}{2}} + \right.
\nonumber\\
&& \left. + \frac{i}{2}(\bar\lambda D\lambda - D\bar\lambda
\lambda)  + \frac{cN \sqrt{k}}{2R} \bar\lambda \lambda +
\frac{\sqrt{3}}{2}c\Lambda^{1/2}N \bar\lambda \lambda + \right.
\label{13}\\
&& \left. + \frac{(-1 + 6\gamma)}{\sqrt{2}} N \tilde G^{1/2}
M_\gamma^{1/2}R^{\frac{-3 - 3 \gamma}{2}}\bar\lambda \lambda +
\frac{c^2 \sqrt{k} R^{1/2}}{2\tilde G^{1/2}}(\bar\psi \lambda -
\psi \bar\lambda)\right.
\nonumber\\
&& \left. + \frac{c^2 \Lambda^{1/2}}{2\sqrt{3}\tilde G^{1/2}}
R^{3/2}(\bar\psi \lambda - \psi \bar\lambda) - \frac{\sqrt{2}}{2}
cM_\gamma^{1/2}R^{-\frac{3\gamma}{2}}(\bar\psi \lambda - \psi
\bar\lambda) \right\}dt,  \nonumber
\end{eqnarray}
where $DR = {\dot R} - \frac{i\tilde G^{1/2}}{2cR^{1/2}}(\psi
\bar\lambda + \bar\psi \lambda)$ and $D\lambda = {\dot \lambda} -
\frac{1}{2} V\lambda$, $D{\bar\lambda} = {\dot {\bar\lambda}} +
\frac{1}{2} V\bar\lambda$.
\begin{center}
{\bf Supersymmetric Quantum Model}
\end{center}
In this section we will proceed with the quantization analysis of
the system. The classical canonical Hamiltonian is calculated in
the usual way for the systems with constraints. It has the form
\begin{equation}
H_c = NH + \frac{1}{2} \bar\psi S - \frac{1}{2}\psi {\bar S} +
\frac{1}{2}VF, \label{14}
\end{equation}
where $H$ is the Hamiltonian of the system, $S$ and ${\bar S}$ are
the supercharges and $F$ is the $U(1)$ rotation generator. The
form of the canonical Hamiltonian (\ref{14}) explains the fact
that $N, \psi, \bar\psi$ and $V$ are Lagrangian multipliers which
only enforce the first-class constraints $H = 0, S = 0, {\bar S }
= 0$ and $F = 0$, which express the invariance under the conformal
$n = 2$ supersymmetric transformations. The first-class
constraints may be obtained from the action (\ref{13}) varying
$N(t),\psi(t)$,$\bar\psi(t)$ and $V(t)$, respectively. The
first-class constraints are
\begin{eqnarray}
H &=& -\frac{\tilde G}{2c^2 R} \pi^2_R - \frac{c^4 k R}{2\tilde G}
- \frac{c^4 \Lambda R^3}{6 \tilde G} - M_{\gamma}c^2R^{-3\gamma} +
\frac{\sqrt{2}c^3\Lambda^{1/2} M_\gamma^{1/2}}{\sqrt{3} \tilde
G^{1/2}} R^{\frac{3 - 3
\gamma}{2}} - \nonumber\\
&-& \frac{c^4 \sqrt{k} \Lambda^{1/2} R^2}{\sqrt{3} \tilde G} +
\frac{\sqrt{2k}c^3}{\tilde G^{1/2}} M_{\gamma}^{1/2}R^{\frac{1 -
3\gamma}{2}} - \frac{c \sqrt{k}}{2R} \bar\lambda \lambda -
\frac{\sqrt{3}}{2}c\Lambda^{1/2} \bar\lambda \lambda -
\nonumber\\
&-& \frac{(6\gamma - 1)}{\sqrt{2}}\tilde G^{1/2}
M_\gamma^{1/2}R^{\frac{-3 - 3 \gamma}{2}}\bar\lambda \lambda
,\label{15}\\
S&=& \Big(\frac{i\tilde G^{1/2}}{c R^{1/2}}\pi_R - \frac{c^2
\sqrt{k} R^{1/2}}{\tilde G^{1/2}} - \frac{c^2 \Lambda^{1/2}
R^{3/2}}{\sqrt{3}\tilde G^{1/2}} + \sqrt{2} c M_{\gamma}^{1/2}
R^{-\frac{3\gamma}{2}}\Big)
\lambda, \label{16} \\
{\bar S}&=& \Big(-\frac{i\tilde G^{1/2}}{c R^{1/2}}\pi_R -
\frac{c^2 \sqrt{k} R^{1/2}}{\tilde G^{1/2}} - \frac{c^2
\Lambda^{1/2} R^{3/2}}{\sqrt{3}\tilde G^{1/2}} + \sqrt{2} c
M_{\gamma}^{1/2} R^{-\frac{3\gamma}{2}}\Big)\bar\lambda, \label{17}\\
F&=& - \bar\lambda \lambda, \label{18}
\end{eqnarray}
where $\pi_R = - \frac{c^2 R}{{\tilde G}N} {\dot R} +
\frac{icR^{1/2}}{2N \tilde G^{1/2}}(\bar\psi \lambda + \psi
\bar\lambda)$ is the canonical momentum associated to $R$. The
canonical Dirac brackets are defined as
\begin{equation}
\lbrace R, \pi_R \rbrace = 1,\quad \lbrace \lambda, \bar\lambda
\rbrace = i. \label{19}
\end{equation}
With respect to these brackets the super-algebra for the
generators $H, S, {\bar S}$ and $F$ becomes
\begin{equation}
\lbrace S, {\bar S}\rbrace = - 2iH,\quad \lbrace S, H \rbrace =
\lbrace {\bar S}, H \rbrace = 0, \quad \lbrace F, S\rbrace =
iS,\quad \lbrace F, {\bar S}\rbrace = i{\bar S}. \label{20}
\end{equation}
In a quantum theory the brackets (\ref{19}) must be replaced by
anticommutators and commutators, they can be considered as
generators of the Clifford algebra. We have
\begin{eqnarray}
\lbrace \lambda, \bar\lambda \rbrace &=& -\hbar, \qquad [R,
\pi_R]= i\hbar \, \qquad {\rm with} \quad \
\pi_R=-i\hbar \frac{\partial}{\partial R} \label{21} \\
\bar \lambda &=&  \xi^{-1} \lambda^{\dag} \xi =- \lambda^{\dag},
\qquad \lbrace \lambda, \lambda^{\dag} \rbrace =  \hbar, \qquad
\lambda^{\dag} \xi= \xi \lambda^{\dag} \quad {\rm and} \quad
\xi^{\dag}=\xi. \nonumber
\end{eqnarray}
Then, for the operator ${\bar S}$ the following equation is
satisfied
\begin{equation}
\bar S=\xi^{-1} S^{\dag} \xi. \label{22}
\end{equation}
Therefore, the anticommutator of supercharges $S$ and their
conjugated operator ${\bar S}$ under our defined conjugation has
the form
\begin{equation}
\overline {\left\{ S, \bar S \right\}}= \xi^{-1} \left\{ S, \bar S
\right\} \xi = \left\{ S, \bar S \right\}, \label{23}
\end{equation}
and the Hamiltonian operator is self-conjugated under the
operation ${\bar H} = \xi^{-1} H^{\dagger} \xi$. We can choose the
matrix representation for the fermionic para\-meters $\lambda,
\bar\lambda$ and $ \xi$ as
\begin{equation}
\lambda = \sqrt{\hbar} \sigma_-, \qquad \bar \lambda = -
\sqrt{\hbar} \sigma_+, \qquad \xi= \sigma_3, \label{24}
\end{equation}
with $\sigma_\pm = \frac{1}{2}(\sigma_1 \pm i\sigma_2)$, where
$\sigma_1$, $\sigma_2$, $\sigma_3$ are the Pauli matrices.
\\

In the quantum level we must consider the nature of the Grassmann
variables $\lambda$ and $\bar\lambda$, with respect to these we
perform the antisymmetrization, then we can write the bilinear
combination in the form of the commutators, $\bar\lambda, \lambda
\to \frac{1}{2}[\bar\lambda, \lambda]$, and this leads to the
following quantum Hamiltonian $H$.
\begin{eqnarray}
H_{quantum} &=& -\frac{\tilde G}{2c^2} R^{-1/2} \pi_R R^{-1/2}
\pi_R - \frac{c^4 k R}{2\tilde G} - \frac{c^4 \Lambda R^3}{6
\tilde G} - M_{\gamma}c^2R^{-3\gamma} \nonumber\\
&+& \frac{\sqrt{2}c^3\Lambda^{1/2} M_\gamma^{1/2}}{\sqrt{3} \tilde
G^{1/2}} R^{\frac{3 - 3 \gamma}{2}} - \frac{c^4 \sqrt{k}
\Lambda^{1/2} R^2}{\sqrt{3} \tilde G} + \nonumber\\
&+& \frac{\sqrt{2k}c^3}{\tilde G^{1/2}} M_{\gamma}^{1/2}R^{\frac{1
- 3\gamma}{2}} - \frac{c \sqrt{k}}{4R}[\bar\lambda, \lambda] -
\frac{\sqrt{3}}{4}c\Lambda^{1/2} [\bar\lambda, \lambda] -
\nonumber\\
&-& \frac{(6\gamma - 1)}{2\sqrt{2}}\tilde G^{1/2}
M_\gamma^{1/2}R^{\frac{-3 - 3 \gamma}{2}}[\bar\lambda, \lambda].
\label{25}
\end{eqnarray}
The supercharges $S$, $\bar S$ and the fermion number $F$ have the
following structures:
\begin{equation}
S = A\lambda, \qquad\qquad S^{\dag} = A^{\dag} \lambda^{\dag}
\label{26}
\end{equation}
where
\begin{equation}
A = \frac{i\tilde G^{1/2}}{c} R^{-1/2} \pi_R - \frac{c^2
\sqrt{k}}{\tilde G^{1/2}} R^{1/2} - \frac{c^2 \Lambda^{1/2}
R^{3/2}}{\sqrt{3}\tilde G^{1/2}} + \sqrt{2}
cM_\gamma^{1/2}R^{-\frac{3\gamma}{2}},\label{27}
\end{equation}
and
\begin{equation}
F = -\frac{1}{2}[\bar\lambda, \lambda]. \label{28}
\end{equation}
An ambiguity exist in the factor ordering of these operators, such
ambiguities always arise, when the operator expression contains
the product of non-commuting operator $R$ and $\pi_R$, as in our
case. It is then necessary to find some criteria to know which
factor ordering should be selected. We propose the following rule
to integrate with the inner product of two states $^{35-38}$. This
inner product is calculated performing the integration with the
measure $R^{1/2} dR$. With this measure the conjugate momentum
$\pi_R$ is non-Hermitian with $\pi^{\dagger}_R = R^{-1/2} \pi_R
R^{1/2}$. However, the combination $(R^{-1/2} \pi_R)^{\dagger} =
\pi_R^{\dagger} R^{-1/2} = R^{-1/2} \pi_R$ is a Hermitian one, and
$(R^{-1/2} \pi_R R^{1/2} \pi_R)^{\dagger} = R^{-1/2}\pi_R
R^{1/2}\pi_R$ is Hermitian too. This choice in our supersymmetric
quantum approach $n = 2$ eliminates the factor ordering ambiguity
by fixing the ordering parameter $p = \frac{1}{2}$
$^{30-34,39,40}$.
\begin{center}
{\bf Superquantum Solutions}
\end{center}
In the quantum theory, the first-class constraints $H = 0, S = 0,
{\bar S} = 0$ and $F = 0$ become conditions on the wave function
$\Psi(R)$. Furthermore, any physical state must be satisfied the
quantum constraints
\begin{equation}
H \Psi(R) = 0, \quad S\Psi(R) = 0,\quad {\bar S}\Psi(R) = 0, \quad
F\Psi(R) = 0, \label{29}
\end{equation}
where the first equation is the Wheeler-DeWitt equation for the
minisuperspace model. The eigenstates of the Hamiltonian
(\ref{25}) have two components in the matrix representation
(\ref{24})
\begin{equation}
\Psi = \pmatrix{\Psi_1\cr \Psi_2 \cr}. \label{30}
\end{equation}
However, the supersymmetric physical states are obtained applying
the supercharges operators $S\Psi = 0, {\bar S}\Psi = 0$. With the
conformal algebra given by (\ref{20}), these are rewritten in the
following form
\begin{equation}
(\lambda {\bar S} - \bar\lambda S)\Psi = 0.\label{31}
\end{equation}
Using the matrix representation for $\lambda$ and $\bar\lambda$ we
obtain the following differential equations for $\Psi_1(R)$ and
$\Psi_2(R)$ components
\begin{eqnarray}
\Big(\frac{\hbar \tilde G^{1/2}}{c} R^{-1/2} \frac{\partial
}{\partial R} - \frac{c^2 \sqrt{k} R^{1/2}}{\tilde G^{1/2}} -
\frac{c^2 \Lambda^{1/2} R^{3/2}}{\sqrt{3} \tilde G^{1/2}} +
\sqrt{2} cM_\gamma^{1/2}R^{-\frac{3\gamma}{2}}
\Big) \Psi_1(R) = 0.\label{32}\\
\Big(\frac{\hbar \tilde G^{1/2}}{c} R^{-1/2} \frac{\partial
}{\partial R} + \frac{c^2 \sqrt{k} R^{1/2}}{\tilde G^{1/2}} +
\frac{c^2 \Lambda^{1/2} R^{3/2}}{\sqrt{3} \tilde G^{1/2}} -
\sqrt{2} cM_\gamma^{1/2}R^{-\frac{3\gamma}{2}} \Big) \Psi_2(R) =
0.\label{33}
\end{eqnarray}
Solving these equation, we have the following wave functions
solutions
\begin{eqnarray}
\Psi_1(R) = C \exp{\Big[ \frac{\sqrt{k} c^{3}R^{2}}{2\hbar \tilde
G} + \frac{c^3 \Lambda^{1/2}}{3 \sqrt{3}\hbar \tilde G}R^{3} -
\frac{2\sqrt{2}c^2M_{\gamma}^{1/2}}{(3 - 3\gamma)\hbar \tilde
G^{1/2}}R^{\frac{3 - 3 \gamma}{2}}\Big]},\\ \label{34} \Psi_2(R) =
\tilde C \exp{\Big[ -\frac{\sqrt{k} c^{3}R^{2}}{2\hbar \tilde G} -
\frac{c^3 \Lambda^{1/2}}{3 \sqrt{3}\hbar \tilde G}R^{3} +
\frac{2\sqrt{2}c^2M_{\gamma}^{1/2}}{(3 - 3\gamma)\hbar \tilde
G^{1/2}}R^{\frac{3 - 3 \gamma}{2}}\Big]}.
\end{eqnarray}
In the case of the flat universe $(k = 0)$ and for the dust-like
matter $(\gamma = 0)$ we have the following solutions (using the
relation $M_{\gamma = 0} = R^3\rho_{\gamma = 0})$
\begin{eqnarray}
\Psi_1(R) = C_1 \exp{\Big[ \frac{1}{\sqrt{6\pi}}\Big(
\frac{\rho_\Lambda}{\rho_{pl}}\Big)^{1/2} \Big(
\frac{R}{l_{pl}}\Big)^3 - \frac{2}{\sqrt{6\pi}} \Big(
\frac{\rho_{\gamma = 0}}{\rho_{pl}}\Big)^{1/2} \Big(
\frac{R}{l_{pl}}}\Big)^3\Big], \label{35}
\end{eqnarray}
\begin{eqnarray}
\Psi_2(R) = C_2 \exp{\Big[ -\frac{1}{\sqrt{6\pi}}\Big(
\frac{\rho_\Lambda}{\rho_{pl}}\Big)^{1/2} \Big(
\frac{R}{l_{pl}}\Big)^3 + \frac{2}{\sqrt{6\pi}} \Big(
\frac{\rho_{\gamma = 0}}{\rho_{pl}}\Big)^{1/2} \Big(
\frac{R}{l_{pl}}}\Big)^3\Big], \label{36}
\end{eqnarray}
where $\rho_{pl} = \frac{c^5}{\hbar G^2}$ is the Planck density
and $l_{pl} = \Big(\frac{\hbar G}{c^3}\Big)^{1/2}$ is the Planck
length.
\\

We can see, that the function $\Psi_1$ in (\ref{35}) has good
behavior when $R \to \infty$ under the condition $\rho_\Lambda < 4
\rho_{\gamma = 0}$, while $\Psi_2$ does not. On the other hand,
the wave function $\Psi_2$ in (\ref{36}) has good behavior when $R
\to \infty$ under the condition $\rho_\Lambda > 4 \rho_{\gamma =
0}$, because the principal contribution comes from the first term
of the exponent, while $\Psi_1$ does not have good behavior.
However, only the scalar product for the second wave function
$\Psi_2$ is normalizable in the measure $R^{1/2}dR$ under the
condition $\rho_\Lambda > 4 \rho_{\gamma = 0}$. This condition
does not contradict the astrophysical observation at
$\rho_{\Lambda} \geq 3 \rho_{M}$, due to the fact that the dust
matter introduces the main contribution to the total energy
density of matter $\rho_M$.
\\

On the other hand, according to recent astrophysical data, our
universe is dominated by a mysterious form of the dark energy
$^{41}$, which counts to about $75-80$ per cent of the total
energy density. As a result, the universe expansion is
accelerating $^{42,43}$. Vacuum energy density $\rho_\Lambda =
\frac{c^2 \Lambda}{8\pi G}$ is a concrete example of the dark
energy.
\begin{center}
{\bf Conclusion}
\end{center}
The recent cosmological data give us the following range for the
dark energy state parameter $\gamma = - 0.96^{+0.08}_{-0.09}$.
However, in the literature we can find different theoretical
models for the dark energy with state parameter $\gamma > - 1$ and
$\gamma < - 1$. In the present work we have discussed the case for
$\gamma = 0$ corresponding to the FRW universe with barotropic
perfect fluid as matter field. In the case of the flat universe $(
k = 0)$ and the dust-like matter $\gamma = 0$ we have obtained two
wave functions. However, only the second wave function is
normalizable under the condition $\rho_\Lambda > 4\rho_{\gamma =
0}$, which leads to the cosmological value $\Lambda
> \frac{32 \pi G}{c^2}\rho_{\gamma = 0}.$
\\

\begin{center}
\noindent{\bf Acknowledgments}\\
\end{center}
We thanks M. Gu\'ia, J. Torres, D.A. Rosales and Carlos Montoro
for several useful remarks.
\\

{\bf References}
\begin{enumerate}
\item{} P.A.M. Dirac, Can. J. Math. 2, 129 (1950).

\item{} P.A.M. Dirac, Lectures on Quantum Mecha\-nics (Academic,
New York, 1965).

\item{} P.A.M. Dirac, Proc. R. Soc. London A 246, 333 (1958).

\item{} P.A.M. Dirac, Phys. Rev. {\bf 114}, 924 (1959).

\item{} R. Arnowitt, S. Deser and C.W. Misner, in Gravitation: An
Introduction to Current Research, edited by L. Witten (Wiley, New
York, 1962).

\item{} J. Anderson, Phys. Rev. {\bf 114}, 1182 (1959).

\item{} J. Anderson, In Proceedings of the First Eastern
Theoretical Physics Conferences, edited by M.E. Rose (Gordon and
Breach, New York, 1963).

\item{} A. Komar, Phys. Rev. {\bf D 20}, 830, (1979).

\item{} C. Teitelboim, Phys. Lett. {\bf B 69}, 240 (1977).

\item{} C. Teitelboim, Phys. Rev. Lett. {\bf 38}, 1106 (1977).

\item{} R. Tabensky and C. Teitelboim, Phys. Lett. {\bf B 69}, 453
(1977).

\item{} A. Mac\'ias, O. Obreg\'on and M.P. Ryan, Jr., Class.
Quantum Grav. {\bf 4}, 1477 (1987).

\item{} P.D. D'Eath and D.I. Hughes, Phys. Lett. {\bf B 214}, 498
(1988).

\item{} P.D. D'Eath and D.I. Hughes, Nucl. Phys. {\bf B378}, 381
(1991).

\item{} J. Socorro, O. Obreg\'on and A. Mac\'ias, Phys. Rev. {\bf
D 45}, 2026 (1992).

\item{} P.D. D'Eath, S.W. Hawking and O. Obreg\'on, Phys. Lett.
{\bf B 300}, 44 (1993).

\item{} P.D. D'Eath, Phys. Rev. {\bf D 48}, 713 (1993).

\item{} M. Asano, M. Tanimoto and N. Noshino, Phys. Lett. {\bf B
314}, 303 (1993).

\item{} R. Capovilla and J. Guven, Class. Quantum Grav. {\bf 11},
1961, (1994).

\item{} P.D. D'Eath, Phys. Lett. {\bf B 320}, 12 (1994).

\item{} A.D.Y. Cheng, P.D. D'Eath and P.R.L.V. Moniz, Phys. Rev.
{\bf D 49}, 5246, (1994).

\item{} R. Capovilla and O. Obreg\'on, Phys. Rev. {\bf D 49},
6562, (1994).

\item{} A. Csord\'as and R. Graham, Phys. Rev. Lett. {\bf 74},
4129 (1994).

\item{} R. Graham, Phys. Rev. Lett. {\bf 67}, 1381, (1991).

\item{} R. Graham, Phys. Lett. {\bf B 277}, 393, (1992).

\item{} J. Bene and R. Graham, Phys. Rev. {\bf D 49}, 799, (1994).

\item{} O. Obreg\'on, J. Socorro and J. Ben\'itez, Phys. Rev. {\bf
D 47}, 4471 (1993).

\item{} O. Obreg\'on, J.J. Rosales and V.I. Tkach, Phys. Rev. {\bf
D 53}, R1750 (1996).

\item{} V.I. Tkach, J.J. Rosales and O. Obreg\'on, Class. Quantum
Grav. {\bf 13}, 2349 (1996).

\item{} C. Ortiz, J.J. Rosales, J. Socorro, J. Torres and V.I.
Tkach. Phys. Lett. {\bf A 340}, 51-58, (2005).

\item{} O. Obreg\'on, J.J. Rosales, J. Socorro and V.I. Tkach,
Class. Quantum Gravity, {\bf 16}, 2861 (1999).

\item{} V.I. Tkach, J.J. Rosales and J. Socorro, Class. Quantum
Gravity, {\bf 16}, 797 (1999).

\item{} M. Ryan Jr, Hamiltonian Cosmology, Springer Verlag (1972).

\item{} J. Socorro, M.A. Reyes and F.A. Gelbert, Phys. Lett. {\bf
A 313}, 338 (2003).

\item{} P.D. D'Eath, D.I. Hughes, Phys. Lett. {\bf B 214}, 498
(1988).

\item{} P.D. D'Eath, D.I. Hughes, Nucl. Phys. {\bf B 378} 381
(1991).

\item{} T. Christodoulakis, J. Zanelli, Phys. Rev. {\bf D 29} 2738
(1984).

\item{} T. Christodoulakis, J. Zanelli, Phys. Lett. {\bf B 102}
237 (1984).

\item{} S.W Hawking, D.N. Page, Nucl. Phys. {\bf B 264} 185
(1986).

\item{} J. Halliwel, in: S. Coleman, et al.(Eds.), Introductory
Lectures on Quantum Cosmology (Jerusalem Winter School), vol. 7,
World Scientific, Singapore, pp. 159-243 1991.

\item{} T. Padmanabhan, Phys. Rept. {\bf 380}, 235 (2003).

\item{} S. Perlmutter, et al; Astrophysics J., {\bf 517}, 565
(1999).

\item{} A.G. Riess, et al; Astrophysics J., {\bf 607}, 665 (2004).
\end{enumerate}


\end{document}